\documentclass[11pt,a4paper]{article}
\usepackage{disitechrep}
\usepackage{times}

        \let\tt=\ttfamily
          
\let\bf=\bfseries

\usepackage{graphicx}
\usepackage{amsfonts}
\usepackage{amssymb}

\begin{document}

\title{Simplification of cross-field topology}

\author
     {Daniele Panozzo and Enrico Puppo\\
        Department of Computer and Information Sciences\\
      University of Genova
   }

\technumber{08}
\date{October 14, 2010}

\maketitle





\section{Statement of the problem}
\label{sec:statement}

Let $M$ be a smooth 2-manifold and let $\cal C$ a smooth cross field defined on $M$.
%
For all general definitions about cross fields, we refer to \cite{RayValLiLev08}.

We assume that $\cal C$ has only a finite number of isolated {\em singularities,} and that all such singularities are of the first order, i.e., the index of a singularity can be either $\frac{1}{4}$, or $-\frac{1}{4}$.
Each point of $M$ that is not a singularity of $\cal C$ will be said to be \emph{regular.}

A {\em streamline} (a.k.a. {\em integral line}) of $\cal C$ is a smooth line on $M$ that is tangent/orthogonal to the directions defined by $\cal C$ at each point;
a streamline either is a cycle, or it has its endpoints at (not necessarily distinct) singularities.
A streamline that ends at singularities is called a \emph{separatrix.}

Let $S$ be the set of singularities of $M$.  
Under the assumptions above, there exist a finite number of separatrices that join pairs of points of $S$, which may cross orthogonally at a finite set $R$ of regular points of $M$.
Separatrices partition $M$ into a finite number of ``quadrangular'' domains, where  by quadrangular, we mean that each region is bounded by four segments of separatrix. 

Let $E$ be the set of segments of separatrix that join two points of $V=S\cup R$. 
Then $G=(V,E)$ is a graph embedded on $M$ with the following properties:
\begin{enumerate}
\item Each minimal cycle is formed by four edges;
\item Each regular vertex in $R$ has valence four;
\item Each singularity in $S$ having index $\frac{1}{4}$ has valence three;
\item Each singularity in $S$ having index $-\frac{1}{4}$ has valence five.
\end{enumerate}

Note that, since $G$ is embedded on $M$, the arcs incident at any vertex $v\in V$ can be arranged in radial order around $v$.
Let $c=(v_0,\ldots,v_n)$ be a chain in $G$, and let $v_i$ be a regular vertex on the chain. 
We say that $c$ \emph{crosses} $v_i$ if the edges $(v_{i-1},v_i)$ and  $(v_i,v_{i+1})$ are not consecutive in the radial order around $v_i$, otherwise we say that $c$ \emph{turns} at $v_i$.
A separatrix corresponds to a chain $(s_0,r_1,\ldots,r_{k-1},s_k)$ where $s_0$ and $s_k$ are singularities, all the $r_i$ are regular and the chain crosses at all regular vertices.

We aim at obtaining another graph $G'=(V',E')$ such that $V'=S\cup R'$, with $R'$ possibly empty, which fulfills all properties 1-4 defined above (with $R'$ in the place of $R$) and such that the lines of $E'$ are \emph{almost aligned} with $\cal C$. 
The problem can be defined as an optimization one, by giving an energy ${\cal E}(G')$ that is proportional to the number of vertices of $R'$ and to the ``warping'' introduced by $G'$ on $\cal C$, which depends on how much the edges of $E'$ deviate from lying on streamlines of the cross field.

In the following section, we remain generic about energy $\cal E$, as well as about geometry of lines of $E$, while we describe a greedy algorithm that provides a heuristic for tackling with this optimization problem in general, by treating it as a graph simplification problem.


\section{Graph simplification}
\label{sec:simpl}

We get in input graph  $G$ as defined in the previous section, and we generate a graph $G'=(S\cup R',E')$ that fulfills properties 1-4, trying to reduce the number of regular vertices in $R'$, while disregarding the geometry of edges in $E'$. 

Our algorithm works through macro-operations that allow us editing the graph while maintaining it consistent with properties 1-4.
Each macro-operation corresponds to a series of atomic operations that may either delete, or detach and reattach edges in the graph, and possibly remove regular vertices. 

Let us consider a vertex $v$ together with its incident edges. 
If we detach an edge from $v$, we generate a \emph{defect} at $v$.
If $v$ is either singular, or regular, then we call this defect either a \emph{S-defect,} or a \emph{R-defect,} respectively. 
If we attach an edge $e$ to a defected vertex $v$, then we repair a defect.
Note that, since edges are arranged radially, each defect has a given position in the radial order about $v$.
Repair is valid only if the position of edge $e$ in the radial order, which depends on the embedding of $e$ on $M$, is coincident with the position of the repaired defect. 
Note that we are interested in such an embedding only just from the point of view of topology, i.e., the shape of $e$ is not relevant, as long as it does not cross other edges. 

\begin{figure}
\centering
\mbox{}
\hfill
\centerline{\includegraphics[width=0.6\linewidth]{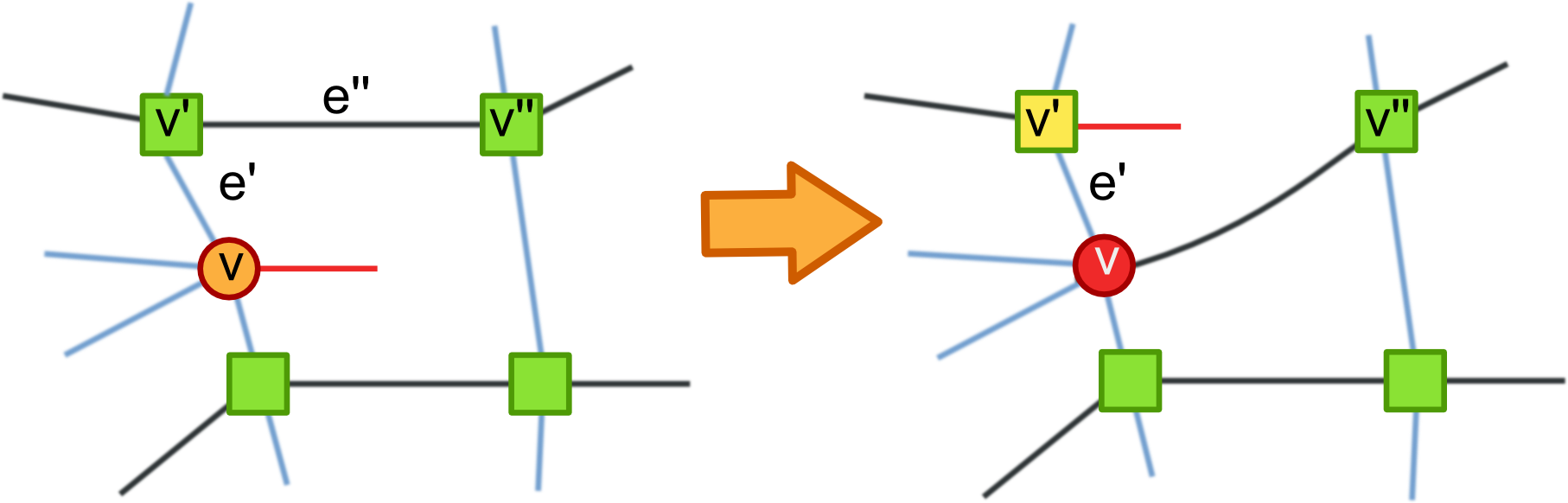}}

\hfill\mbox{}
\caption{\label{fig:repair} Sub-atomic operation used to repair a S-defect. Circles and squares represent singular and regular vertices; defected vertices are depicted in orange and yellow; defects are represented as red segments; the defect at vertex $v$ can be repaired by cutting one of the two separatrices depicted in black.
}
\end{figure}

Repairing a defect corresponds to the sub-atomic operation described in the following. 
In our algorithm, there can be just one defect per vertex (actually, there will be just one S-defect and one R-defect in the whole graph after each atomic operation). 
A defect at vertex $v$ is repaired by cutting one of the two separatrices that run ``parallel'' to the defected edge, and attaching one of the two branches of the cut separatrix to $v$ at the defect.
This is accomplished as depicted in Figure \ref{fig:repair}: let $e'$ the edge adjacent to the defect, in the radial counterclockwise order about $v$, and let $v'$ the other endpoint of $e'$; if $v'$ is regular, let $e''$ the edge next to $e'$ in counterclockwise order about $v'$, and let $v''$ be the other endpoint of $e''$; edge $e''$ is deleted and $v$ is connected to $v''$ with a new edge; this generates a R-defect at $v'$, and the branch of cut separatrix ending at such defect will be called the \emph{dangling} branch.

The same operation could be done symmetrically by rotating clockwise about $v$ and cutting the other ``parallel'' separatrix, therefore there are two possible options for repairing a defect. 
If a neighboring vertex $v'$ is singular, the corresponding option cannot be used to repair the defect; 
if both neighboring vertices are singular, then the defect cannot be repaired, and the algorithm will have to backtrack (see next about backtracking).

\begin{figure}
\centering
\mbox{}
\hfill
\includegraphics[width=1\linewidth]{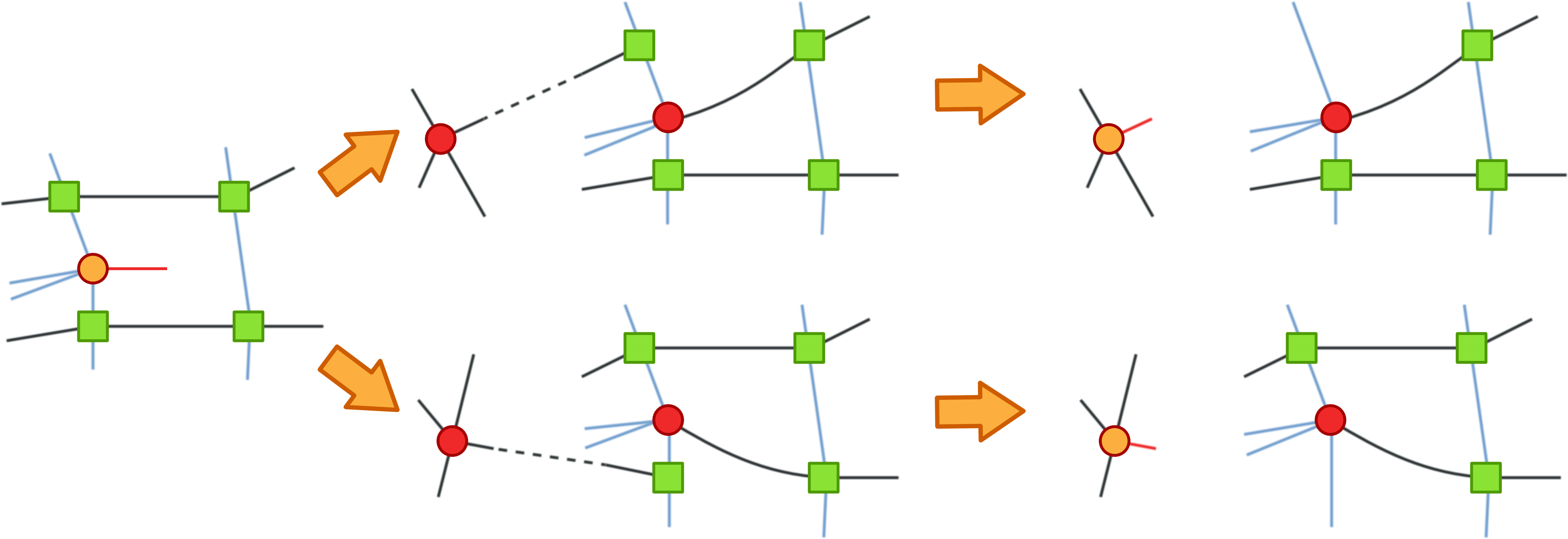}

\hfill\mbox{}
\caption{\label{fig:op2} The two possible configurations generated by a Switch-separatrix operation.
}
\end{figure}

Atomic operations are of the following two types:
\begin{enumerate}
\item \textbf{Delete-separatrix} (Figure \ref{fig:op}): delete all edges and regular vertices of a separatrix; for each regular vertex $v$ deleted, the two other edges incident at $v$ are merged to form a single edge; one of the two S-defects caused by removing the separatrix is repaired (as explained above), by generating a R-defect; 

\begin{figure}
\centering
\mbox{}
\hfill
\centerline{\includegraphics[width=0.8\linewidth]{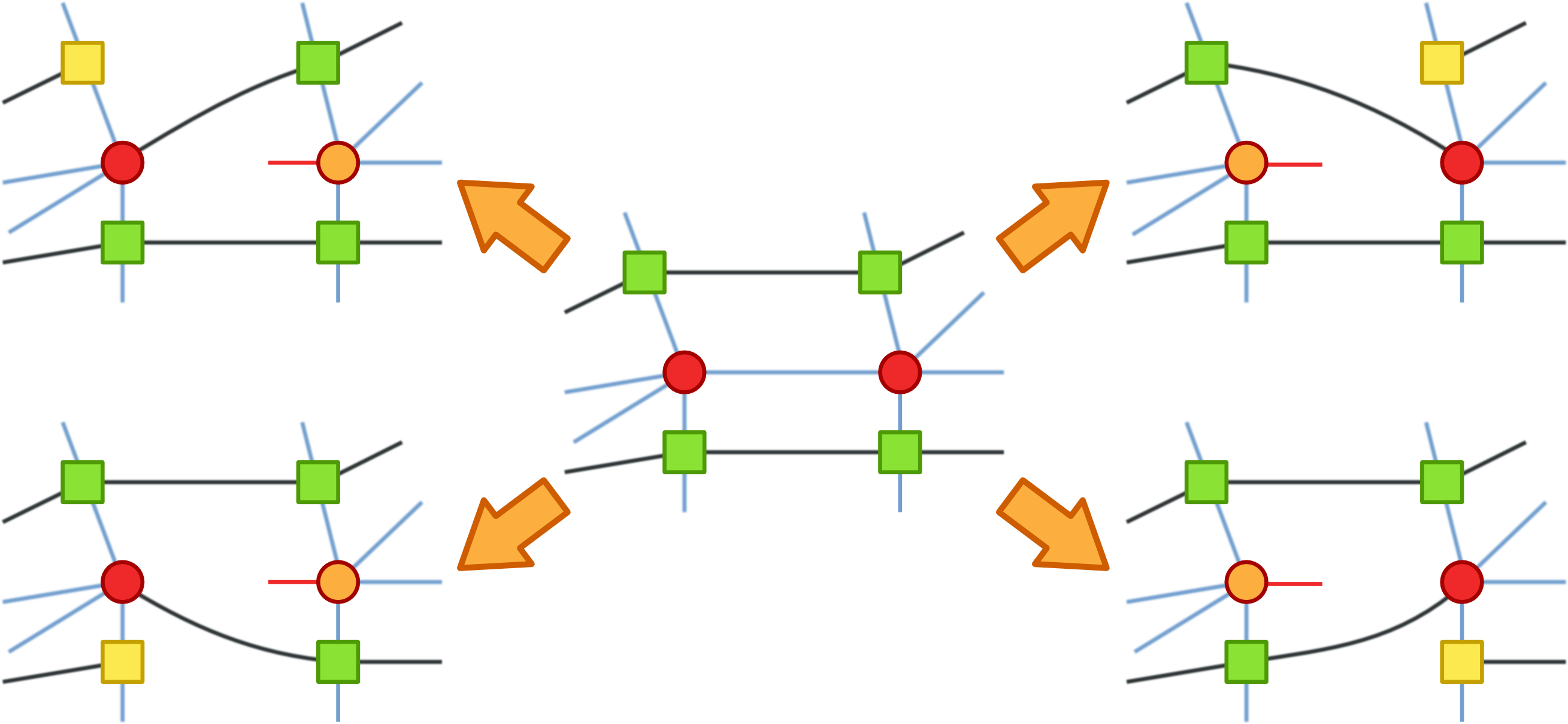}}

\hfill\mbox{}
\caption{\label{fig:op} The four possible configurations generated by the \textbf{Delete-separatrix} operation.
}
\end{figure}

\item \textbf{Switch separarix} (Figure \ref{fig:op2}): repair a R-defect by cutting a separatrix (as explained above) and delete the dangling branch of the separatrix; delete all regular vertices traversed by the dangling branch and merge pairs of edges incident at them, as in the previous operation; if the other end of the dangling branch is a singular vertex, then a S-defect is generated.
\end{enumerate}

A macro-operation consists of one \textbf{Delete-separatrix} operation followed by a loop of \textbf{Switch-separatrix} operations, which ends when no S-defect remains in the graph.
Note that after each operation, one S-defect and one R-defect are present in the whole graph. 
Each \textbf{Switch-separatrix} removes a S-defect and potentially produce another S-defect; however, if the dangling branch of separatrix removed by this operation ends at the (only) R-defect, then no defects are left. 
This sets the stop condition for the cycle in a macro-operation.

Now let us consider the possible choices we have in applying a single macro-operation. 
Let $n$ be the total number of separatrices of field $\cal C$ (note that this number remains constant throughout processing).
We thus have $n$ options to give a separatrix in input to \textbf{Delete-separatrix}, and four options to implement this operation, as shown in Figure \ref{fig:op}, hence a total of $4n$ options to initialize the simplification process.   
Given one such choice, once we enter the loop, we have two options for each \textbf{Switch-separatrix} operation, as shown in Figure \ref{fig:op2}.
This gives a total space of solutions consisting of a forest of $4n$ binary trees, one for each initial choice.
Leaves of a tree correspond to either termination a solution, or a stuck condition, which occurs when a defect cannot be repaired.
All trees are guaranteed to have a finite number of nodes, because each \textbf{Switch-separatrix} operation removes at least one regular vertex and there exist a finite number of such vertices.

An algorithm searching for the best simplification (with respect to some given criterion) could explore this space of solutions and select the best one (i.e., the one that minimized the given energy {\cal E}).
The complexity is combinatorial, but it could be feasible in practice, because separatrices are usually not very numerous, and the depth of trees is also usually quite small.

Our approach aims at evaluating a suboptimal solution with a greedy algorithm that explores just a path in a tree from the forest, possibly backtracking and choosing an alternative path if it gets stuck at a leaf. 
The separatrix to be deleted and the way to repair the first S-defect are selected with some heuristic; at each cycle, the way to repair the R-defect is selected with some other heuristic. 
The algorithm keeps track of the operations performed in a stack. 
If it gets stuck at a leaf, it backtracks by removing the last operation from the stack, and by freezing the corresponding option, until a node with an open (non-frozen) child is found.
In the worst case, the whole tree is visited while no path leads to a solution, so the graph is left unchanged.

The repeated application of a macro-operation simplifies the graph progressively. 
This process can be repeated until some used-defined criterion occurs, such as having reached either a given number/percentage of regular vertices, or a maximum tolerated level of warping with respect to field $\cal C$.

\section{Optimization criteria}
\label{sec:optimize}
We now consider the geometric aspects involved in the simplification process, in order to devise some heuristics for driving our algorithm.
We need a criterion to choose the first separatrix and S-defect to be repaired, and a criterion to chose between the two possible options to repair a given defect.

\begin{figure}
\centering
\mbox{}
\hfill
\centerline{\includegraphics[width=0.6\linewidth]{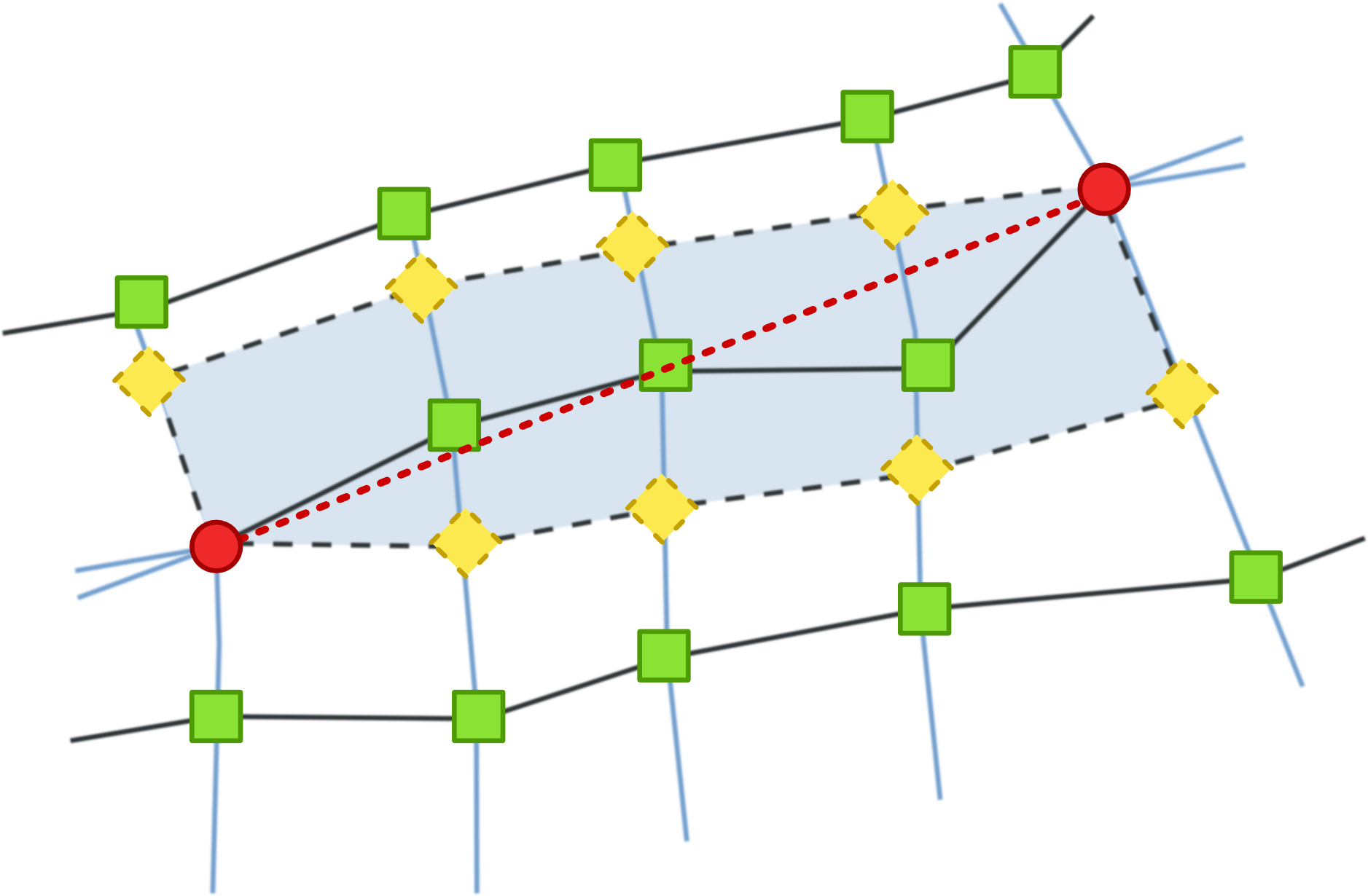}}

\hfill\mbox{}
\caption{\label{fig:drift} The minimal drift is the ration between the area of highlighted in blue and the square of the length of the dotted diagonal. The yellow square denotes regular vertexes in the graph that has been removed by simplification operators.
}
\end{figure}

Since our aim is to reduce the number of regular vertices of $G$, we start the algorithm by deleting the separatrix that crosses the largest number of vertices. 
The criterion for selecting the edge to detach to repair a defect is the same in both atomic operations and it is based on the concept of \emph{minimal drift} explained in the following and depicted in Figure \ref{fig:drift}. 

Let us consider a defect at a vertex $v$, and let $v'$, $e'$ and $e''$ be described as in the previous section.
Let $s_0$ be the separatrix leaving $v$ in the direction of the defect and let $s_0^{\prime}$ the separatrix leaving $v$ in the direction of $e'$, both separatrices being computed on the original graph $G$.
Similarly, let  $s_1$ be the branch of separatrix through $v'$ in the direction of $e''$, let $v_s$ be the singularity at the end of such branch, and let $s_1^{\prime}$ be the separatrix leaving $v_s$ in the direction orthogonal to $s_1$.
The four separatrices $s_0$,   $s_0^{\prime}$, $s_1$ and $s_1^{\prime}$ form a quadrangular region $Q$ on $M$.
Let $A_Q$ be the area of $Q$, and let $\tilde{s}$ be the shortest line through $Q$ joining $v$ to $v_s$
Finally, let $l_{\tilde{s}}$ be the length of $\tilde{s}$.
We define the {\em drift} of $\tilde{s}$ as the ratio between the area of $Q$ and the square length of $\tilde{s}$,
$$\frac{A_Q}{l_{\tilde{s}}^2}.$$ 
The drift is roughly proportional to the ratio between the ``thickness'' and the ``length'' of strip $Q$ and it roughly measures how much line $\tilde{s}$ deviates from being a streamline of $\cal C$. 

Assuming that we can measure this drift efficiently, we choose each time to repair a defect by using the edge corresponding to the line with minimum drift. 

\section{Implementation}
\label{sec:implementation}

Our prototype implementation works on cross-fields induced by quadrangular meshes.
Let $M$ be a quadrangular mesh, such that all its vertices have valencies between three and five.
Mesh $M$ stands here as a discrete approximation of a smooth manifold.
We may assume a (discrete approximation of) a cross field $\cal C$ to be defined on $M$ by assuming that:
\begin{itemize}
\item The edges of $M$ are aligned with the streamlines of $\cal C$;
\item $\cal C$ is regular inside each face of $M$.
\end{itemize}
Under these assumptions, the singlularities of $\cal C$ correspond to the irregular vertices of $M$, and separatrices correspond to chains of edges that start at those irregular vertices and cross at all regular vertices.
The input graph $G$ is obtained easily by traversing each separatrix and tagging all its edges and vertices.
Thus, $G$ will consist of all tagged vertices and edges of $M$ after traversing all separatrices.

In Figure \ref{fig:fert}, we show the graph of separatrices obtained from the quadrangulation of a popular dataset, obtained with the method presented in \cite{BomZimKob09}.
Each separatrix is depicted with a different color.
Note how some separatrices may take involved routes. 
In particular, some of them form long helices, which generate many intersections with other separatrices, thus many regular vertices of $G$.
These separatrices will be the first ones to be selected for deletion by our algorithm.

If the input mesh $M$ is highly regular, as in the case of meshes obtained with \cite{BomZimKob09}, the algorithm can be greatly simplified by assuming that all edges of $M$ have the same (unit) length.
Note that each edge of $G$ can consist of a chain of edges of $M$, thus it will have an integer length.
Under these assumptions it is trivial to compute both the length of a separatrix and the drift, as described in the previous section. 

Once the graph has been simplified, we actually compute each new separatrix $\tilde{s}$ as depicted in Figure \ref{fig:drift}, by assuming it as a ``diagonal'' of strip $Q$:
in fact, strip $Q$ is a regular rectangular sub-grid of $M$, with the endpoints of  $\tilde{s}$ at two opposite corners of the grid. 

\section{Results}
\label{sec:results}

\begin{figure}
\centering
\includegraphics[width=0.3\linewidth]{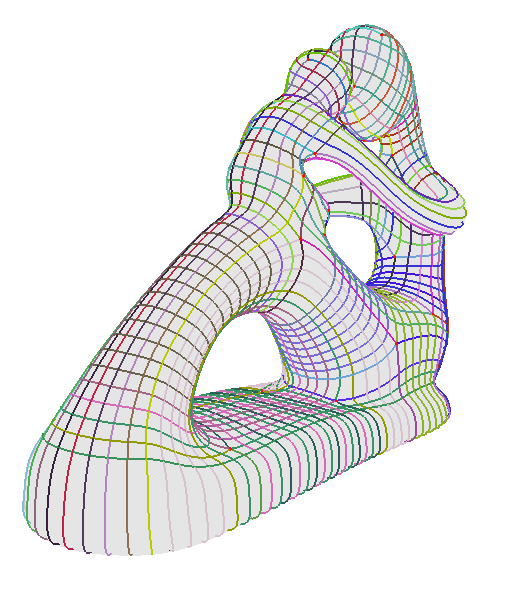} 
\includegraphics[width=0.3\linewidth]{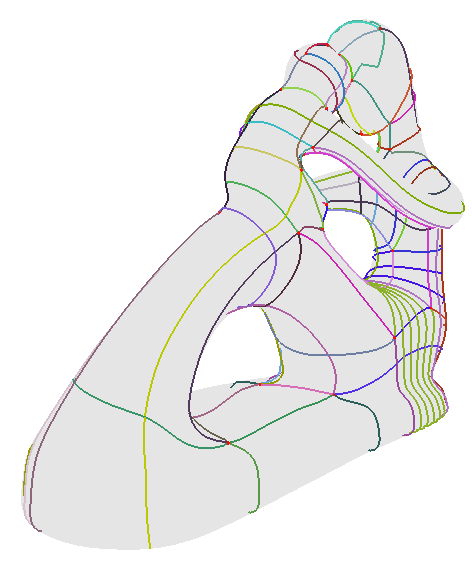} 
\includegraphics[width=0.3\linewidth]{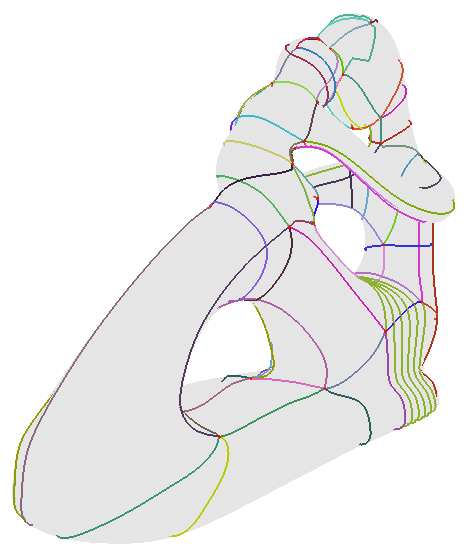} 
\hspace{1.5cm} (a) \hspace{4cm} (b) \hspace{4cm} (c) 

\hfill\mbox{}
\caption{\label{fig:fert} The graph of separatrices computed on Fertility model remeshed with \cite{BomZimKob09} (a). The graph simplified with a single step of simplification (b) . A second simplification step reduces the number of regular vertices with respect to the original model of about 96\% (c).
}
\end{figure}

The first experimental results obtained with our method are depicted in Figure \ref{fig:fert}.
The input mesh \emph{Fertility} consists of 3357 quads and it has been obtained with the remeshing method in \cite{BomZimKob09}.
The mesh contains 48 singularities.
The graph $G$ of separatrices induced by such mesh contains 108 separatrices consisting of 2217 regular vertices.
Among them, there are some long helices, which generate many intersections with other separatrices, thus many regular vertices of $G$.
These separatrices will be the first ones to be selected for deletion by our algorithm.

Figure \ref{fig:fert} (a) show the graphs of separatrices of the original mesh. After one step of simplification the graph contains only 197 regular vertices and it shown in Figure \ref{fig:fert} (b). A second simplification step allows to reduce the number of regular vertices to 88. It is not possible to further simplify the mesh without introducing a very high drift.

Since our algorithms minimize the drift during the simplification, singularities are naturally aligned as it can be seen in the base or in the back of the mother. 

\section{Concluding remarks}
\label{sec:conc}
This draft paper presents just preliminary work.
Our method can work equally on line fields containing singularities of the first order of type \emph{lemon} and \emph{star.}
We plan to extend our method to work on line and cross fields computed on generic meshes. 
There exist many methods for computing such fields, most of which work on triangle meshes.
A challenging problem is to extract graph $G$ of such fields without computing a quadrangular remeshing aligned with the cross field.






\bibliographystyle{disitechrep}
\bibliography{extra}

\end{document}